\documentclass[letterpaper, 10 pt, conference]{ieeeconf}  

\IEEEoverridecommandlockouts                              
\usepackage[utf8]{inputenc}
\usepackage[T1]{fontenc}
\usepackage{graphicx}
\graphicspath{ {./images/} }

\usepackage{sgame}
\usepackage{amssymb}
\usepackage{amsmath}
\usepackage{xcolor}
\usepackage{float}
\usepackage{comment}
\usepackage{lipsum}
\usepackage{mathbbol}
\usepackage[capitalize]{cleveref}
\usepackage{enumerate}
\usepackage{cite}
\usepackage{mathtools}
\usepackage{tikz}

\usetikzlibrary{decorations.pathreplacing}

\usepackage{cuted}
\newtheorem{remark}{Remark}
\newtheorem{proposition}{Proposition}
\newtheorem{theorem}{Theorem}

\newtheorem{definition}{Definition}

\newtheorem{lemma}{Lemma}

\DeclareMathOperator{\Equaldef}{\overset{def}{=}}

\title{\LARGE \bf
Strategic information disclosure with communication constraints and private preferences}

\author{Marcos M. Vasconcelos and Odilon C\^{a}mara    
\thanks{M. M. Vasconcelos is with the Department of Electrical and Computer Engineering, FAMU-FSU College of Engineering, Florida State University,  Tallahassee, FL, 32306, USA. O. Camara is with the Department of Finance and Business Economics, Marshall School of Business, University of Southern California, Los Angeles, CA, 90089, USA. E-mails:
        {\tt m.vasconcelos@fsu.edu, odilon.camara@marshall.usc.edu}.}%
}

\begin{document}

\maketitle
\thispagestyle{empty}
\pagestyle{empty}

\begin{abstract}

Social-media platforms are one of the most prevalent communication media today. In such systems, a large amount of content is generated and available to the platform. However, not all content can be transmitted to every possible user at all times. At the other end are the users, who have their own preferences about which content they enjoy, which is often unknown ex ante to the platform. We model the interaction between the platform and the users as a signaling game with asymmetric information, where each user optimizes its preference disclosure policy, and the platform optimizes its information disclosure policy. We provide structural as well as existence of policies that constitute Bayesian Nash Equilibria, and necessary optimality conditions used to explicitly compute the optimal policies.


\end{abstract}

\section{Introduction}
Most today's information is disseminated and consumed through social-media platforms \cite{allon2019information}. This paradigm shift allows platforms to actively curate the content they show or promote to their user base, generating more revenue by better targeting the right audience \cite{candogan2020optimal}. Additionally, most platforms allow users to express their preferences for the type of information they are interested in consuming. Traditionally, these preferences are collected in discrete forms, such as ``like,'' ``dislike,'' and ``indifference,'' for example. In this paper, we model this setting as a signaling game with asymmetric information, where both the platform (transmitter) and the user (receiver) aim to minimize the same loss function. Due to the asymmetry of information, their expected losses lead to different optimization problems, resulting in the emergence of a Bayesian game. Our goal is to characterize the equilibrium policies for this game and provide a framework that enables their computation.

Strategic information transmission has a rich history, \textcolor{black}{ including models of communication of ``hard'' information \cite{milgrom1981good} and cheap-talk models \cite{crawford1982strategic}.}
Many variants 
have emerged with applications in both Economics \cite{Camara:2024} and Engineering \cite{Akyol:2017}. Within Engineering, the field of remote estimation has developed, where sensors make strategic decisions about whether to communicate their observations to a receiver \cite{farokhi2016estimation}. The receiver's goal is to form real-time estimates of a random state of the system based on the selectively disclosed information.



\begin{figure}[t!]
    \centering
\includegraphics[width=0.8\columnwidth]{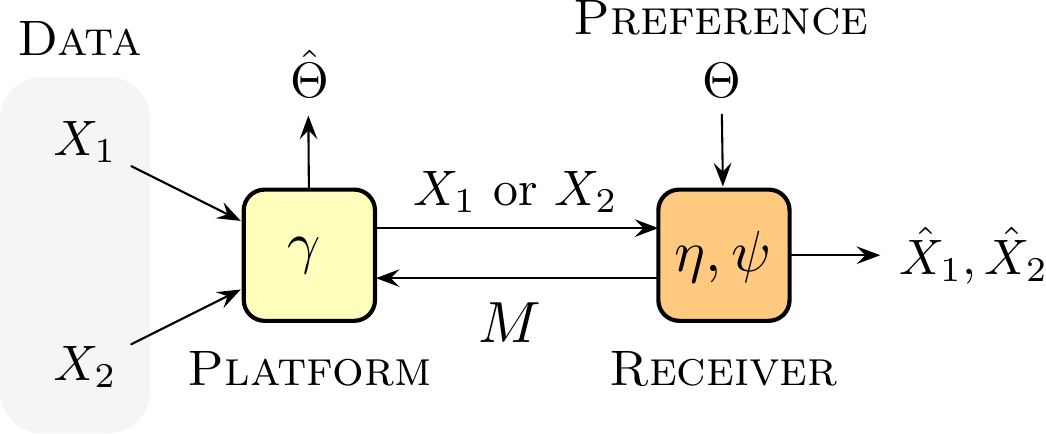}
    \caption{Schematic diagram for the strategic information transmission problem in the presence of private receiver preferences.}
    \label{fig:diagram}
\end{figure}

We consider the setting depicted in \cref{fig:diagram}, where two hard-information sources are available to the Platform, which selects one to disclose to the Receiver. The Receiver forms estimates of both sources but has a random private preference for one of them. To enable the Platform to select the correct information source, the Receiver sends a cheap-talk message about its preference to the Platform. However, due to a communication constraint, the Receiver can only send a finite number of bits, requiring the Platform to estimate the Receiver's preference in order to adjust its information disclosure policy. In classic strategic disclosure games, imperfect communication typically arises because Sender and Receiver have different preferences. We consider the case in which they have the same preference but face constraints in their communication channels, that is, Platform and User can only exchange a limited amount of information.

The setting considered here is based on the work of Vasconcelos and Mitra \cite{Vasconcelos:2020}, which introduced the basic setting of disclosing one out of two correlated Gaussian random variables to a receiver, which is equally interested in both. Later, the model was extended to account for power constraints at the transmitter \cite{Vasconcelos:2020b}, lack of knowledge about the data's prior distributions \cite{Vasconcelos:2021}, and the use of neural-network approximations to unveil properties of the often elusive globally optimal scheduling policies \cite{vasconcelos2024neuroscheduling}.   

In our model, the goal of the platform is aligned with the users in the sense that both the user and the platform are working together to minimize the expected loss function. However, the asymmetry in the information pattern and the communication constraints lead to a non-trivial problem\textcolor{black}{: the platform knows the realized values of the two variables but does not know their relative importance, while the opposite is true for the receiver.} The presence of a \textit{private preference} is the main innovation in the model presented herein. Our main contribution is to establish the existence of a structured pair of disclosure policies for the platform and the receiver that constitute a Bayesian Nash Equilibrium. From the point of view of the receiver, it uses a monotone partition of the interval of possible preferences akin to a task-driven quantization \cite{gray2006quantization} to send a cheap-talk message to the platform, which we call Monotone Preference Disclosure Policy. At the opposite end, the Platform, uses a policy that compares weighted quadratic errors, where the weights are computed as a conditional expectation of the receiver's preference, \textcolor{black}{with the Platform updating its belief based on} the cheap-talk message \textcolor{black}{received} and the knowledge about the \textcolor{black}{User's} preference disclosure policy. 

\textcolor{black}{A similar set of results were derived by C\^amara, Matsusaka, and Chong in \cite{Camara:2024} for a very different model of communication, in which multiple senders can privately acquire information. The senders compete to sell a binary (approve/not approve) recommendation to receivers who have publicly known preferences and must cast a vote to approve or not approve a proposal. Despite the extreme differences between their model and ours, they also find that, in equilibrium, receivers are divided into partitions and recommendations are tailored to the average receiver in each partition.}

Our work is related to the growing literature on information disclosure \cite{Velicheti:2023,velicheti2023value,Velicheti:2024} and Bayesian persuasion and its variants \cite{ALONSO2016672,Massicot:2023}. Overall, signaling games \cite{saritacs2016quadratic} continue to play a fundamental role in understanding interactions between strategic agents and can help predict behavior in social networks in the presence of misinformation \cite{Hebbar:2022}. 

The rest of the paper is organized as follows. In Section II, we present the problem setup, state the optimization problem, assumptions, and adopted solution concept. In Section III, we present our main theoretical results, which consist of a structural characterization of a broad class of equilibria for the game, and its existence. In Section IV, we define the value of privacy and present a numerical example to support our theoretical results. The paper concludes in Section V, where several directions for future work are suggested.

\section{Problem setup}

Consider an asymmetric signaling game between a Transmitter (Platform) and a Receiver (User) communicating according to the block diagram in \cref{fig:diagram}. The transmitter has access to two pieces of content, represented by two random variables $X_1$ and $X_2$.  Let $X_1$ and $X_2$ be independent random variables, distributed according to probability density functions $f_{X_1}(x_1)$ and $f_{X_2}(x_2)$, which are symmetric around their means $\mu_1$ and $\mu_2$, and have full support on the real numbers. The receiver has a random preference $\Theta$, \textcolor{black}{which represents the relative weight that the receiver attaches to $X_1$ versus $X_2$, that is, the receiver's preference bias towards these content sources}. We assume that $\Theta$ is \textcolor{black}{independent of $X_1$ and $X_2$,} supported on the unit interval $[0,1]$, and is distributed according to a prior $\pi(\theta)$. \textcolor{black}{All distributions are smooth (no atoms).}

\subsection{Communication constraints}


Let \( n \) be a finite integer denoting the cardinality of the alphabet the user employs to transmit a message to the platform about its private preference. Let \( [n] = \{1, \ldots, n\} \) and \( m \) denote the message from the user to the platform such that \( m \in [n] \). The channel between the user and the platform is a finite-rate, discrete, noiseless channel with a capacity of \( \log_2 n \) bits per channel use. This constraint can model the fact that most platforms limit users to providing information in the form of ``like,'' ``neutral,'' and ``dislike,'' for example.

The communication constraint from the point of view of the platform is a limit on what it is able to send to the user in the number of pieces of content. Here, we assume the simplest case in which the platform has two pieces of content ($x_1,x_2$), but can only send one of the them to the user. Notice that, under this constraint, the platform can transmit the realization of one continuous random variable perfectly, but not both.



\subsection{Game sequence}

The variables $X_1,X_2$  and $\Theta$, are realized. The platform has access to $X_1=x_1,X_2=x_2$, and the user has access to $\Theta=\theta$.

The user acts first, by using a  preference disclosure policy, $\psi:[0,1]\rightarrow [n].$ According to this policy, the user selects a message $m$ as follows:
\begin{equation}
m = \psi (\theta),
\end{equation}
and sends it to the platform. 

Then, the platform uses an information disclosure policy $\gamma:[n] \times\mathbb{R}^2\rightarrow \{1,2\}$, to decide which information source will be made available to the user. Let $s$ denote the platform's decision
, which is computed according to:
\begin{equation}
    s = \gamma(m,x_1,x_2).
\end{equation}



Once a decision is made, i.e., $s$ is chosen, then a signal $y$ is sent to the receiver. Conditioned on $X_1=x_1$ and $X_2=x_2$, the signal $y \in \mathcal{Y}(x_1,x_2)$, where 
\begin{equation}
\mathcal{Y}(x_1,x_2) \Equaldef \big\{(1,x_1), (2,x_2)\big\}.
\end{equation}
Given $X_1=x_1$ and $X_2=x_2$, the signal $y$ is determined as follows:
\begin{equation}
    y = \begin{cases}
    (1,x_1) & \text{if} \ \ s = 1 \\
    (2,x_2) & \text{if} \ \ s = 2.
    \end{cases}
\end{equation}

The receiver, upon observing the signal $y$, outputs two estimates, one for each content source, $\hat{x}_i\in \mathbb{R},$ $i\in\{1,2\},$ which are determined using an estimation policy $\eta: \textcolor{black}{\mathcal{Y}(x_1,x_2)\times[n]}\rightarrow \mathbb{R}^2$. 


To estimate the variables, the receiver makes inferences based on the Platform's disclosure strategy, which is a function of the message $m$, rather than the true preference $\theta$, which the Platform cannot directly observe.
In other words, the receiver uses both the sent message and the received signal to output the estimates for the content sources.
That is,
\begin{equation}
(\hat{x}_1,\hat{x}_2) = \eta(y,
m)=\begin{cases}
\big(x_1,\mathbf{E}[X_2|m,y]\big) &   \text{if} \  y=(1,x_1)  \\
  \big(\mathbf{E}[X_1|m,y],x_2\big) &   \text{if} \ y=(2,x_2).
\end{cases}
\end{equation}
In equilibrium, this expectation must be computed by taking into account the equilibrium strategies and applying Bayes's rule.

\subsection{Agent's Loss function}

Given $X_1=x_1,X_2=x_2$ and $\Theta = \theta$, we assume the same following weighted quadratic loss function for the transmitter and the receiver: 
\begin{equation}\label{eq:distortion}
\ell(x_1,x_2,\theta,\hat{x}_1,\hat{x}_2) \Equaldef \theta(x_1-\hat{x}_1)^2 +(1-\theta)(x_2-\hat{x}_2)^2.
\end{equation}
The loss function captures the fact that a user is interested in both variables but he assigns relative weight $\theta$ to variable $x_1$ and $1-\theta$ to variable $x_2$. Hence, a user with a high $\theta$ cares relative more about correctly estimating variable $x_1$.

Due to the asymmetry in the information pattern of the problem, the receiver observes $\Theta = \theta$ but not $X_1 = x_1$ and $X_2 = x_2$. Therefore, the receiver must choose a message $m$ that minimizes its expected loss function:
\begin{equation}
\mathcal{J}_R(m \mid \theta,\gamma,\eta) \Equaldef  \mathbf{E}\Big[\ell(X_1,X_2,\Theta,\hat{X}_1,\hat{X}_2) \mid \Theta = \theta\Big], 
\end{equation}
where the expectation is taken with respect to the receiver's private information about $\theta$, its belief about the platform's information disclosure strategy $\gamma$, and the estimation function $\eta$.

The platform, on the other hand, observes \(X_1 = x_1\), \(X_2 = x_2\), and $M=m$ but not \(\Theta = \theta\). Therefore, the platform must choose the disclosure decision \(s\) that minimizes the following expected loss function:
\begin{multline}
\mathcal{J}_P(s \mid m,x_1,x_2,\psi,\eta) \Equaldef \\ 
\mathbf{E}\Big[\ell(X_1,X_2,\Theta,\hat{X}_1,\hat{X}_2) \mid X_1=x_1,X_2=x_2,M=m \Big],
\end{multline}
where the expectation takes into account the platform's private information about \((x_1,x_2)\), its belief about the receiver's preference disclosure strategy \(\psi\), the receiver's message \(m\), and the estimation function \(\eta\).




\vspace{5pt}

\subsection{Solution concept} 

\begin{definition}
The triplet $(\psi^\star,\gamma^\star,\eta^\star)$ constitutes a \textit{Bayesian Nash Equilibrium} (BNE) if the following conditions hold:
\begin{enumerate}
    \item The following inequality holds for all $\theta\in\Theta$ and $m\in[n]$:
    \begin{equation}
\mathcal{J}_R(m^\star \mid \theta,\gamma^\star,\eta^\star)\leq\mathcal{J}_R(m \mid \theta,\gamma^\star,\eta^\star),
    \end{equation}
    where $m^\star=\psi^\star(\theta)$.
    \item The following inequality holds for all $m\in[n]$, $(x_1,x_2)\in\mathbb{R}^2$, and $s\in\{1,2\}$: 
    \begin{equation}
  \mathcal{J}_P(s^\star \mid  m,x_1,x_2,\psi^\star,\eta^\star)\leq\mathcal{J}_P(s \mid m,x_1,x_2,\psi^\star,\eta^\star),
    \end{equation}
    where $s^\star=\gamma^\star(m,x_1,x_2)$.
    \item The estimation function $\eta^\star$ is consistent with agents' policies and the information available to the receiver.
\end{enumerate}
\end{definition}

\vspace{5pt}

\begin{definition}
We say that the platform uses a \textit{weighted-quadratic disclosure} (WQD) policy if
\begin{equation}
    \gamma(m,x_1,x_2) = \begin{cases}
    1, & \text{if} \ \ w_m(x_1-\mu_1)^2\\ & \ \ \ \ \ \ \ \ \ \ >(1-w_m)(x_2-\mu_2)^2 \\ 
    2, & \text{otherwise},
    \end{cases}
\end{equation}
where $w_m\in[0,1]$ is the weight that the platform assigns to variable $X_1$ when the receiver selects the message $m\in[n]$. 
\end{definition}

\begin{figure}[t!]
    \centering
\includegraphics[width=0.7\columnwidth]{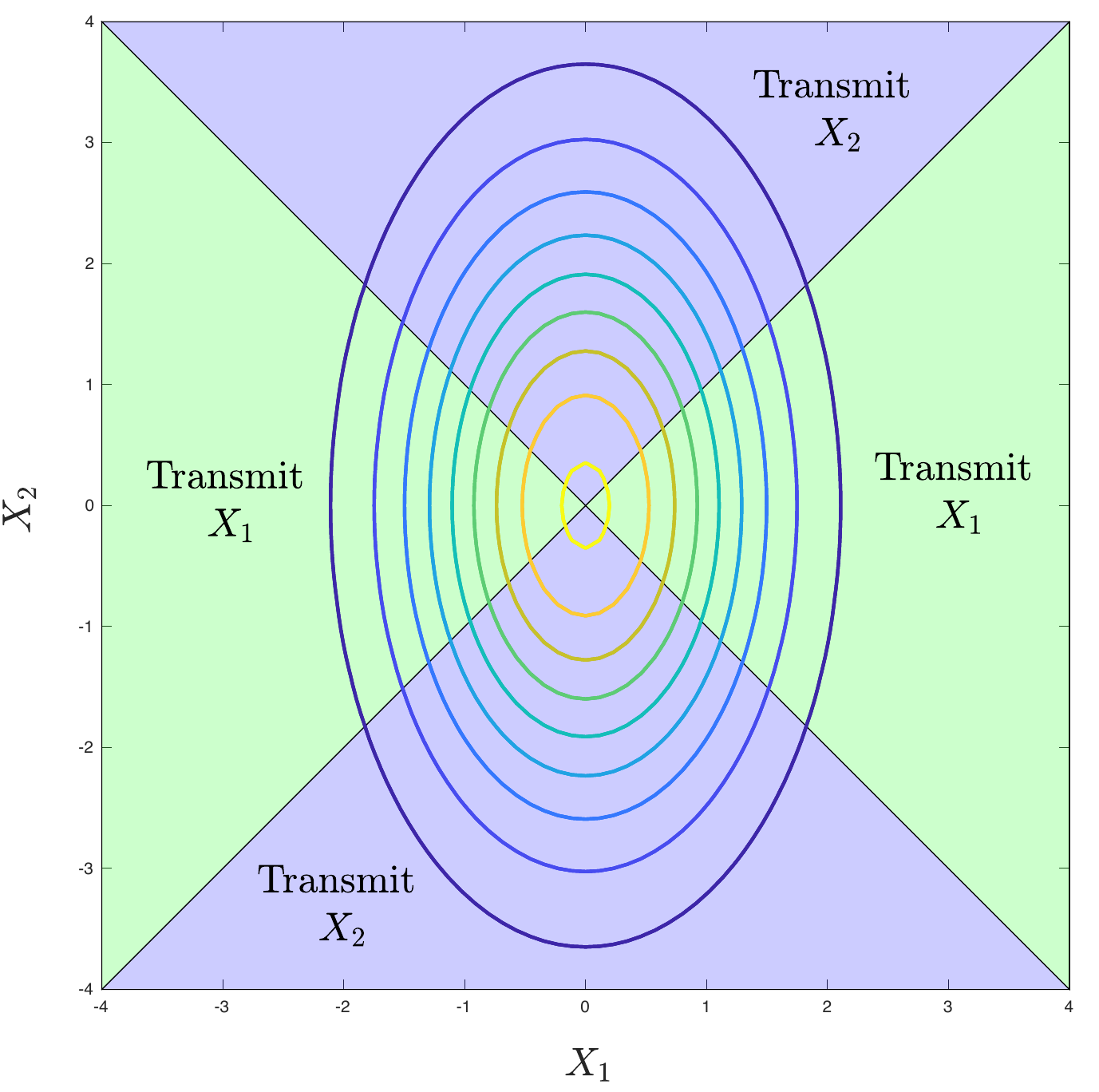}
    \caption{Weighted Quadratic Disclosure Policy with $w_m=1/2$. The ellipsoids correspond to the level curves of a zero-mean bivariate Gaussian probability density with independent components with variances 1 and 2, respectively. }
    \label{fig:WQDP}
\end{figure}

\vspace{5pt}

\begin{remark}
Note that $(x_i-\mu_i)^2$ is the quadratic deviation from the mean. If $w_m=1/2$ then the sender equally weights the two variables and discloses the variable with the largest quadratic deviation from its mean. \Cref{fig:WQDP} shows an example of a WQD policy with parameter $w_m=1/2.$ When $\theta>1/2$, the sender attaches higher weight to variable $X_1$, and will only disclose variable $X_2$ if its quadratic deviation from the mean is sufficiently higher. In the extreme case when $\theta=1$, the platform will always disclose variable $X_1$. 
\end{remark}

\vspace{5pt}



\section{Main Result}

Our main result is to show that there exists a BNE in which the sender uses a weighted-quadratic disclosure policy, and fully characterize its associated preference disclosure policy. We start by providing some insights on the preferences of the receivers. Suppose that the platform uses a WQD policy parameterized by the vector $\bar{w}=(w_1,\ldots,w_n)$. Without loss of generality, we assume that the weights are such that $w_k\leq w_{k+1}$, $k\in[n-1]$. 



\vspace{5pt}

\begin{proposition}
\label{prop.eta}
If the platform uses a WQD policy, $\gamma(x_1,x_2,m)$, the equilibrium estimation policy $\eta^\star$ is:
 \begin{equation}
 \label{eq.etastar}
    \eta^\star(y,m) \Equaldef \begin{cases}
    (x_1,\mu_1), & \text{if} \ \ y=(1,x_1) \\ 
    (\mu_2,x_2), & \text{if} \ \ y=(2,x_2).
    \end{cases}
\end{equation}
\end{proposition}

\vspace{5pt}

\begin{proof} 
If $y=(i,x_i),$ then,
\begin{IEEEeqnarray}{rCl}
\mathbf{E}\big[X_i\mid M=m,Y=(i,x_i)\big]  &=& x_i. \end{IEEEeqnarray}

On the other hand, if $y=(j,x_j),$ $j\neq i$, then
\begin{multline}
\mathbf{E}\big[X_i\mid Y=(j,x_j)\big]  \\ = \int_\mathbb{R}x_i f_{X_i \mid M=m, S=j,X_j=x_j}(x_i)dx_i.
\end{multline}
Since $X_i \perp \!\!\! \perp X_j$, we have
\begin{multline}
f_{X_i \mid M=m,S=j,X_j=x_j}(x_i) \\ = \frac{\mathbf{P}(S=j \mid M=m,X_j=x_j,X_i=x_i)f_{X_i}(x_i)}{\mathbf{P}(S=j \mid X_j=x_j)}.
\end{multline}
Without loss of generality, let $j=1,i=2$. Then,
\begin{multline}
\mathbf{P}(S=1 \mid M=m,X_1=x_1,X_2=x_2) \\= \mathbf{1}\Big((1-w_m)(x_2-\mu_2)^2 < w_m(x_1-\mu_1)^2\Big),
\end{multline}
which implies that
\begin{multline}
\int_{\mathbb{R}}x_2\mathbf{1}\Big((1-w_m)(x_2-\mu_2)^2 < w_m(x_1-\mu_1)^2\Big)f_{X_2}(x_2)dx_2  \\
= \int_{\mu_2-\sqrt{\frac{w_m}{1-w_m}}|x_1-\mu_1|}^{\mu_2+\sqrt{\frac{w_m}{1-w_m}}|x_1-\mu_1|}x_2f_{X_2}(x_2)dx_2.
\end{multline}
Given that $f_{X_2}(x_2)$ is a symmetric distribution around $\mu_2$, we have:
\begin{equation}
\mathbf{E}\big[X_2\mid M=m,Y=(1,x_1)\big] = \mu_2, \ \ x_1\in \mathbb{R}.
\end{equation}
A similar conclusion is reached for $j=2$, and $i=1.$
\end{proof}

\vspace{5pt}

Having established the structure of the equilibrium estimator if the platform uses a WQD policy, the receiver's expected loss function becomes:
\begin{multline}\label{eq:Loss}
    L(w_m \mid \theta) \Equaldef
    {\theta}\mathbf{E}\Big[X_1^2\mathbf{1}\big((1-w_m)X_2^2 \geq w_mX_1^2 \big)\Big]\\
    +(1-{\theta})\mathbf{E}\Big[X_2^2\mathbf{1}\big((1-w_m)X_2^2 <w_m X_1^2 \big)\Big].
    \end{multline}

\vspace{5pt}

\subsection{Quasi-convexity of the Expected Loss Function}

Before stating and proving our main result, we must establish the quasi-convexity property of the expected loss function $L(w_m \mid \theta)$ in \cref{eq:Loss}. Define the following functions:
\begin{equation}
H(w_m) \Equaldef \mathbf{E}\Big[X_2^2 \mathbf{1}\big((1-w_m)X_2^2<w_mX_1^2\big)\Big]
\end{equation}
and 
\begin{equation}
G(w_m) \Equaldef \mathbf{E}\Big[X_1^2 \mathbf{1}\big((1-w_m)X_2^2 \geq w_mX_1^2\big)\Big].
\end{equation}

\vspace{5pt}

\begin{lemma}Let $X_1$ and $X_2$ be independent continuous random variables distributed according to densities $f_{X_1}$ and $f_{X_2}$ symmetric around their respective means $\mu_1$ and $\mu_2$. The following identity holds:
\begin{equation}
\label{eq.Lemma1}
H'(w_m) = - \left(\frac{w_m}{1-w_m}\right) G'(w_m),
\end{equation}
where $H'$ and $G'$ denote the derivatives of $H$ and $G$ with respect to $w_m$. Moreover, $H$ is strictly monotone increasing and $G$ is strictly monotone decreasing in $w_m$.
\end{lemma}

\vspace{5pt}

\begin{proof}
Without loss of generality, assume $\mu_1=\mu_2=0$. The derivative of $H$ with respect to $w_m$ is:
\begin{equation}
H(w_m)= \mathbf{E} \Bigg[ \underbrace{\int_{-\sqrt{\frac{w_m}{1-w_m}}|X_1|}^{\sqrt{\frac{w_m}{1-w_m}}|X_1|} x_2^2f_{X_2}(x_2) dx_2}_{\Equaldef \circledast}   \Bigg].
\end{equation}
Then,
\begin{equation}
\frac{d\circledast}{dw_m} = \frac{w_m^{1/2}}{(1-w_m)^{5/2}}|X_1|^3f_{X_2}\Bigg( \sqrt{\frac{w_m}{1-w_m}}|X_1|\Bigg),
\end{equation}
which implies $H^\prime(w_m)>0$. Similarly, we have:
\begin{equation}
G(w_m)= \mathrm{Var}(X_1) - \mathbf{E} \Bigg[ X_1^2 \underbrace{\int_{-\sqrt{\frac{w_m}{1-w_m}}|X_1|}^{\sqrt{\frac{w_m}{1-w_m}}|X_1|} f_{X_2}(x_2) dx_2}_{\Equaldef \circledast \! \circledast}   \Bigg].
\end{equation}
Then,
\begin{equation}
\frac{d\circledast \! \! \circledast}{dw_m} = \frac{1}{w_m^{1/2}(1-w_m)^{3/2}}|X_1|^3f_{X_2}\Bigg( \sqrt{\frac{w_m}{1-w_m}}|X_1|\Bigg),
\end{equation}
which implies $G^\prime(w_m)<0$. Result \cref{eq.Lemma1} follows from algebraic manipulation of $H'$ and $G'$.
\end{proof}

\vspace{5pt}

\begin{figure}[t!]
    \centering
\includegraphics[width=0.8\columnwidth]{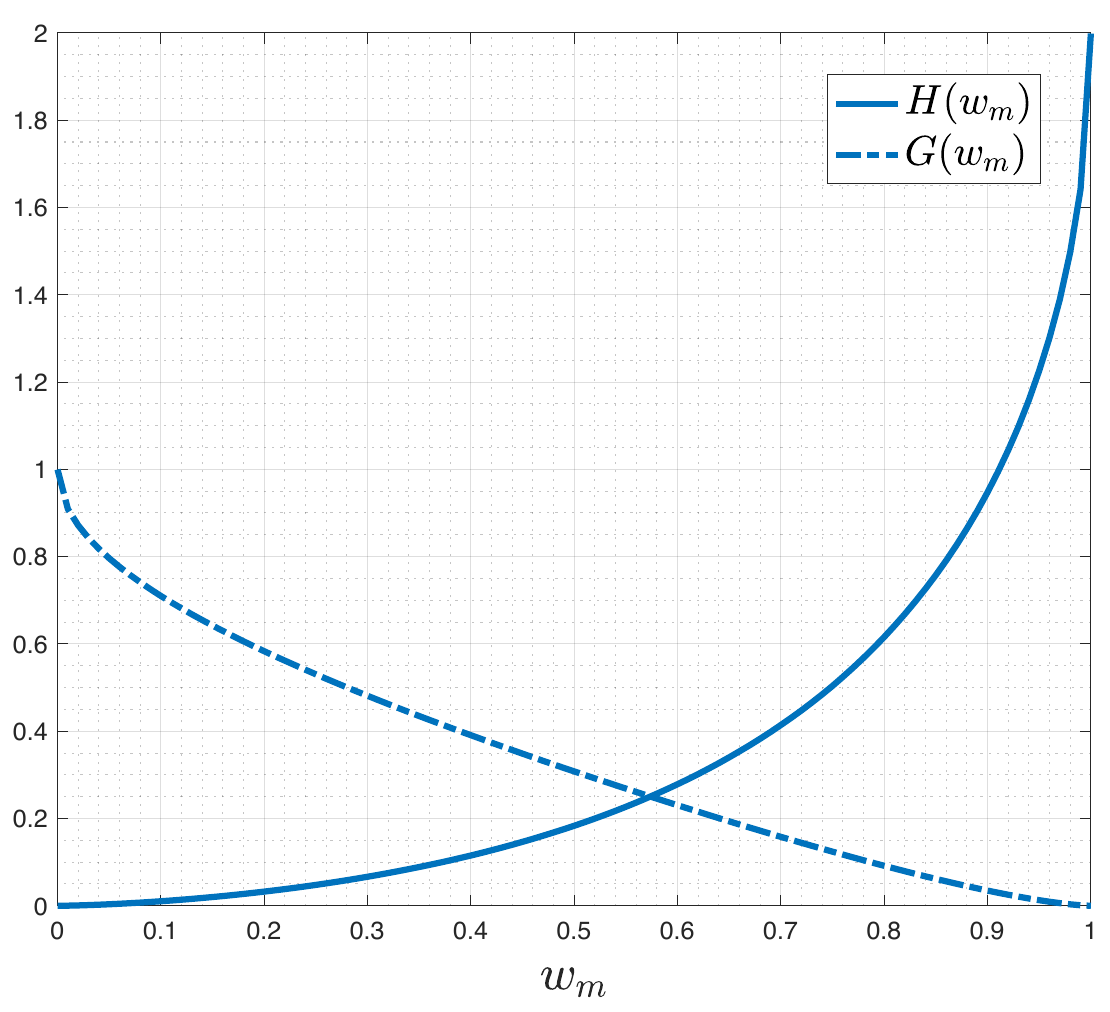}
    \caption{Monotonicity of $H$ and $G$ for $X_1\sim\mathcal{N}(0,1)$ and $X_2\sim\mathcal{N}(0,2)$.}
    \label{fig:Monotonicity}
\end{figure}

\Cref{fig:Monotonicity} displays the strict monotonicity property of $H$ and $G$ for $X_1\sim \mathcal{N}(0,1)$ and $X_2\sim \mathcal{N}(0,2)$. 

\vspace{5pt}

\begin{lemma}
For any symmetric probability density functions $f_{X_1}$ and $f_{X_2}$, the function $L(w_m \mid \theta)$ is strictly quasi-convex in $w_m$, and is minimized at $w_m ^\star = \theta$.
\end{lemma}

\vspace{5pt}

\begin{proof}
Computing the partial derivative of $L(w_m \mid \theta)$ with respect to $w_m$ on the open interval $(0,1)$, we have:
\begin{IEEEeqnarray}{rCl}
\frac{\partial L(w_m \mid \theta)}{\partial w_m}  & = & \theta G'(w_m) + (1-\theta)H'(w_m) \nonumber \\ 
& \stackrel{(a)}{=} & \Bigg(\theta-(1-\theta)  \left(\frac{w_m}{1-w_m}\right)\Bigg) G'(w_m) \nonumber  \\
& = &  \left( \frac{\theta-w_m}{1-w_m}\right)G'(w_m),
\end{IEEEeqnarray}
where $(a)$ follows from \cref{eq.Lemma1}.


From the strict monotone decreasing property of $G$ established in Lemma 1, $G'(w_m)<0$ for  $w_m\in (0,1)$. From the expression for the partial derivative of $L(w_m \mid \theta)$, we have:
\begin{equation}
w_m < \theta \Rightarrow\frac{\partial L(w_m \mid \theta)}{\partial w_m}  < 0; 
\end{equation}
\begin{equation}
w_m > \theta \Leftrightarrow\frac{\partial L(w_m \mid \theta)}{\partial w_m}  > 0; 
\end{equation}
and 
\begin{equation}
w_m = \theta \Leftrightarrow\frac{\partial L(w_m \mid \theta)}{\partial \hat{\theta}}  = 0. 
\end{equation}
\end{proof}

\begin{figure}[t!]
    \centering
    \includegraphics[width=0.8\linewidth]{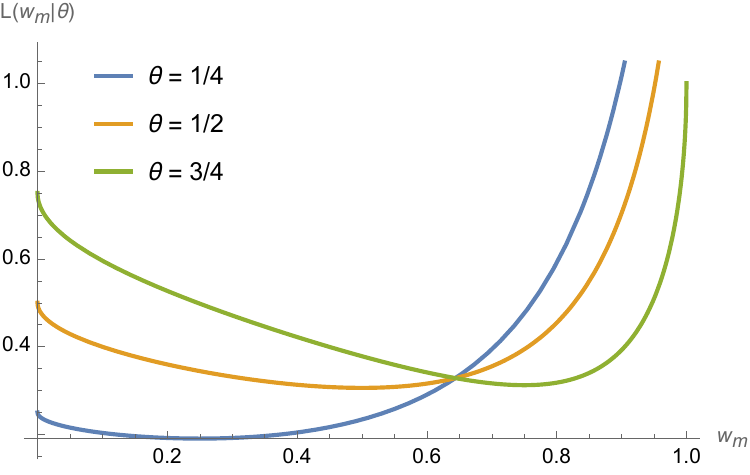}
    \caption{Loss function when $X_1\sim\mathcal{N}(0,1)$ and $X_2\sim\mathcal{N}(0,2)$, for different values of $\theta$: blue line $\theta=1/4$, yellow line $\theta=1/2$, and green line $\theta=3/4$.}
    \label{fig:loss}
\end{figure}

\vspace{5pt}

\begin{definition}
We say that a policy $\psi$ for the receiver is a \textit{Monotone Partition Preference Disclosure} (MPPD) policy if there are $n+1$ thresholds $\{\theta^\star_{0,1},\ldots,\theta^\star_{n,n+1}\}$, with $\theta^\star_{m-1,m}<\theta^\star_{m,m+1}$, and $\theta^\star_{0,1}=0$, $\theta^\star_{n,n+1}=1$ such that:  
\begin{equation}
\psi(\theta)=m, \ \ \text{for all} \ \  \theta \in(\theta^\star_{m-1,m},\theta^\star_{m,m+1}). 
\end{equation}
\end{definition}

\vspace{5pt}

Figure \ref{fig:partitions} illustrates an MPPD with $n$ messages. Within each partition, all receivers send the same message. Notice that messages with larger value are sent by receivers with larger values of $\theta$.

\subsection{Structural Results}

\vspace{5pt}

\begin{lemma}\label{lem:receiver} Suppose that the platform uses a WQD policy with weights $0<w_1<\cdots<w_m<1$. 
Then the \textit{receiver's best response} is an MPPD strategy. 
The thresholds $\theta^\star_{m,m+1}$ are the solutions of the following equation: 
\begin{equation}\label{eq:thresholds}
L(w_m \mid \theta^\star_{m,m+1})=L(w_{m+1} \mid \theta^\star_{m,m+1})
\end{equation}
    for $1\leq m<n$, and we define the corners $\theta_{0,1}\Equaldef0$ and $\theta_{n,n+1}\Equaldef 1$. 
\end{lemma}

\vspace{5pt}

\begin{proof}
When the Platform uses a WQD policy $\gamma^\star$, the estimation policy $\eta^\star$ is given by Eq. \eqref{eq.etastar} and the Receiver's decision problem becomes
\begin{equation}
\min_{m \in \{1,\ldots,n\}} \ \ L(w_m \mid \theta).
\end{equation}

\begin{figure}[t!]
    \centering
    \includegraphics[width=0.8\linewidth]{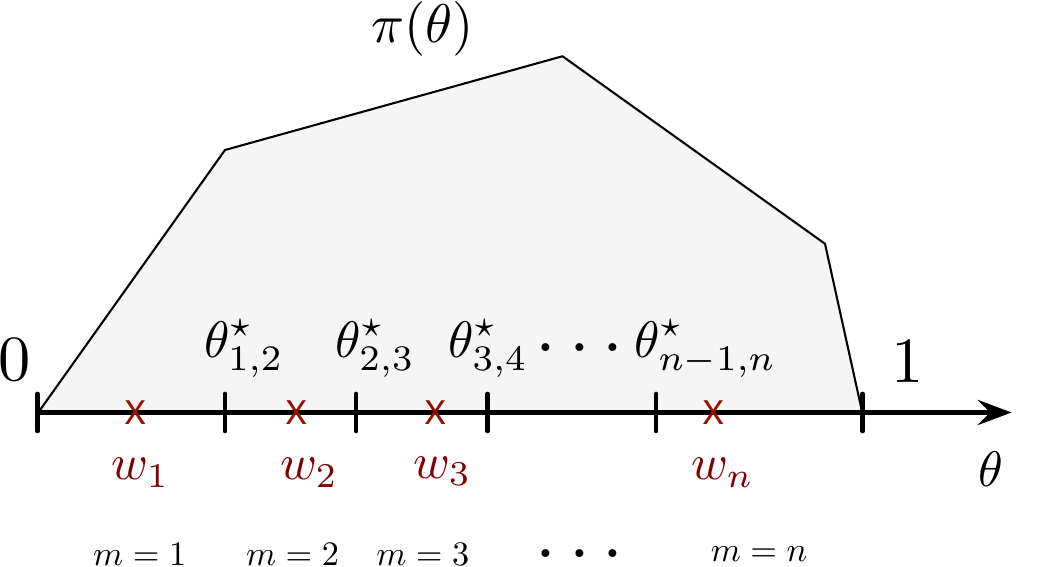}
    \caption{Illustration of a Monotone Partition Preference Disclosure (MPPD) policy.}
    \label{fig:partitions}
\end{figure}

Since this is an optimization over a finite set, the user needs to compare the $n$ possible values of $L(w_m \mid \theta)$. However, since $L(w_m \mid \theta)$ is quasi-convex in $w_m$ with a minimum in $\theta$, it just needs to compare two such values. The receiver proceeds as follows: 
For any $\theta \in [0,1]$, there is a unique index $\tilde{m}\in [n]$ such that $w_{\tilde{m}} \leq \theta < w_{\tilde{m}+1}$.

Due to the quasi-convexity of $L(w_m \mid \theta)$, we have:
\begin{equation}
L(w_{\tilde{m}} \mid \theta) \leq L(w_m \mid \theta), \ \ m< \tilde{m}.
\end{equation}
Similarly, 
\begin{equation}
L(w_{\tilde{m}+1} \mid \theta) \leq L(w_m \mid \theta), \ \ m> \tilde{m}+1.
\end{equation}

Therefore, to compute the decision of what message to send to the platform about its private variable, the optimal decision rule is given by:
\begin{equation}\label{eq:dec_rule}
L(w_{\tilde{m}+1} \mid \theta)  \underset{m^\star=\tilde{m}}{\overset{m^\star=\tilde{m}+1}{\gtrless}} L(\tilde{m} \mid \theta).
\end{equation}
\Cref{eq:dec_rule} is characterized by a single threshold $\theta^\star_{\tilde{m},\tilde{m}+1}$ denoting the value of the preference where the user is indifferent to choosing $m^\star=\tilde{m}$ or  $m^\star=\tilde{m}+1$, and is given by \cref{eq:thresholds}. Notice that the threshold $\theta^\star_{\tilde{m},\tilde{m}+1}$ always exists and is unique. However, we omit the proof of this technical result for brevity.
\end{proof}

\vspace{5pt}

\begin{lemma}
\label{lem:sender}
Suppose that the receiver uses the estimation policy \textcolor{black}{$\eta^\star$ described by \cref{eq.etastar}.}
Let $\mathbf{E}[\Theta\mid M=m,\psi^\star]$ be the sender's belief about the receiver's expected type after observing message $m$, \textcolor{black}{taking into account the receiver's policy $\psi^\star$}. Then it is optimal for the sender to use a WQD policy with weights $w_m=\mathbf{E}[\Theta \mid M=m,\psi^\star]$ for each $m$. That is, the sender chooses the optimal weight from the point of view of the average receiver who sends message $m$.
\end{lemma}

\vspace{5pt}

\begin{proof}
Fix the receiver's \textcolor{black}{ estimation policy $\eta^\star$ to be \cref{eq.etastar}}, and consider the sender's best response, i.e., the value of $s$ that minimizes 
\begin{multline}\label{eq:J_T}
\mathcal{J}_P(s \mid m,x_1,x_2,\psi^\star,\eta^\star) \Equaldef \mathbf{E}\Big[\Theta(X_1 - \hat{X}_1)^2 \\+ (1 - \Theta)(X_2 - \hat{X}_2)^2 \mid M=m, X_1=x_1, X_2=x_2,\psi^\star,\eta^\star\Big].
\end{multline}

The expectation in \cref{eq:J_T} can be expressed as:
\begin{multline}
\mathcal{J}_P(s \mid m,x_1,x_2,\psi^\star,\eta^\star)\\ = \begin{cases}
\big(1-\mathbf{E}[\Theta \mid M=m,\psi^\star]\big)(x_2-\mu_2)^2, & s=1 \\
\mathbf{E}[\Theta \mid M=m,\psi^\star](x_1-\mu_1)^2, & s=2.
\end{cases}
\end{multline}

Let $w_m\Equaldef \mathbf{E}[\Theta \mid M=m,\psi^\star]$, then the  transmission policy that minimizes the expected loss is given by:
\begin{equation}
\gamma^\star(m,x_1,x_2) = 1 \Leftrightarrow (1-w_m) (x_2-\mu_2)^2 < w_m (x_1-\mu_1)^2,  
\end{equation}
and 
\begin{equation}
\gamma^\star(m,x_1,x_2) = 2 \Leftrightarrow (1-w_m) (x_2-\mu_2)^2 \geq w_m (x_1-\mu_1)^2,  
\end{equation}

\end{proof}

\vspace{5pt}

\begin{theorem}[Structure]Consider the asymmetric information disclosure game with \textcolor{black}{disclosure policies $\psi$ and $\gamma$, and estimation policy $\eta$.} Suppose that:
    \begin{enumerate}[A.]
        \item The receiver uses a MDPP policy $\psi^\star$ parameterized by $(\theta^\star_{1,2},\hdots,\theta^\star_{n-1,n})$, with  $\theta^\star_{0,1}=0$ and $\theta^\star_{n,n+1}=1$; 
        \item The platform uses a monotone WQD policy $\gamma^\star$ parameterized by $(w^\star_1,\hdots,w^\star_n)$;  
        \item And the receiver uses \textcolor{black}{the estimation policy $\eta^\star$ described by \cref{eq.etastar}.}
        \end{enumerate}
        The policies $\psi^\star,\eta^\star,\gamma^\star$ constitute a BNE if the following conditions hold:
        \begin{enumerate}[C1.]
        \item For each interior partition threshold,
           \begin{equation}
           \label{eq.C1}
           L(w^\star_m \mid \theta^\star_{m,m+1})=L(w^\star_{m+1}\mid \theta^\star_{m,m+1}).
           \end{equation}
        \item For each $m\in [n]$,
       \begin{equation}
       \label{eq.C2}
w^\star_m=\mathbf{E}\big[\Theta \mid \theta^\star_{m-1,m}<\Theta<\theta^\star_{m,m+1}\big].
    \end{equation}
        \end{enumerate}
\end{theorem}

\vspace{5pt}

\begin{proof}
Given $\eta^\star$, we know from \cref{lem:sender} that the sender's optimal policy must be such that: \begin{equation}
w_m=\mathbf{E}[\Theta \mid M=m,\psi^\star].
\end{equation}
The receiver's monotone partition policy implies that: 
\begin{equation}
\mathbf{E}[\Theta \mid M=m,\psi^\star]=E\big[\Theta \mid \theta^\star_{m-1,m}<\Theta<\theta^\star_{m,m+1}\big],
\end{equation}
therefore the sender is playing a best response. Given $\eta^\star$ and the sender's monotone WQD policy, \Cref{lem:receiver} implies that the receiver's best response must be a monotone partition characterized by the thresholds $\theta^\star_{m,m+1}$ satisfying \begin{equation}
L(w^\star_m \mid \theta^*_{m,m+1})=L(w^\star_{m+1}|\theta^\star_{m,m+1}).
\end{equation}
Hence, these strategies form a BNE if conditions C1 and C2 hold.
\end{proof}

\subsection{Existence Result}

To establish the existence of a BNE for our problem, we will look at the design of an optimal \textcolor{black}{WQD} policy for a \textit{social planner}, which optimizes the weights $w_1,\ldots,w_n$ with the goal of minimizing the expected loss of the entire distribution of receivers.

\vspace{5pt}

\begin{theorem}[Existence]
Suppose that a social planner could choose the weight vector $\bar{w}=(w_1,\ldots,w_n)$ 
\textcolor{black}{ of the WQD policy} that minimizes the \textit{ex ante} expected loss of the receiver. Then, there exists a BNE in which the platform uses these weights.
\end{theorem}

\vspace{5pt}

\begin{proof}
  Given a weight vector $\bar{w}$, the \textit{ex ante} expected loss of the receiver is
    \begin{equation}
    \mathcal{L}(\bar{w})\equiv\int_0^1\min\big\{L(w_1 \mid \theta),\ldots,L(w_n\mid \theta)\big\}\pi(\theta)d\theta.
    \end{equation}
    This is true because a receiver with preference $\theta$ will choose the message $m$ such that $L(w_m \mid \theta)\leq L(w \mid\theta)$ for all $m$.
    Since each function $L(w_m\mid\theta)$ is continuous in $\bar{w}$ and the minimum of continuous functions is continuous, we have that $\mathcal{L}(\bar{w})$ is continuous. The vector $\bar{w}$ belongs to the compact set $[0,1]^n$. Therefore, from the Extreme Value Theorem, there exists $\bar{w}^\star\in[0,1]^n$ that minimizes $\mathcal{L}$.

    We proceed by defining the first-order optimality conditions that must hold at the minima. First, it must be the case that $w_i\neq w_j$ for all $i\neq j$. By contradiction, if there is a pair of equal weights $w_i= w_j$, then one can strictly decrease $\mathcal{L}(\bar{w})$ by changing $w_j$ to any other number $w^\prime\in(0,1)$ that is not equal to another weight in $\bar{w}$. This change strictly decreases the loss of the receiver when his preference is in the neighborhood $w^\prime$, while keeping constant the loss of all other receiver types. 

    Therefore, without loss, we can reorder the weights such that there is an optimal $\bar{w}$ such that $w_1<\hdots<w_n$. Using the results from \cref{lem:receiver}, we can rewrite the expected loss function as 
    \begin{multline}
    \mathcal{L}(\bar{w})=\int_0^{\theta^\star_{1,2}}L(w_1 \mid \theta)\pi(\theta)d\theta+\cdots\\ +    \int_{\theta^\star_{n,n+1}}^1L(w_n \mid\textbf{}\theta)\pi(\theta)d\theta,
    \end{multline}
    where $\theta^\star_{m,m+1}$ is defined as the unique $\theta^\star_{m,m+1}\in[w_m,w_{m+1}]$ such that
    \begin{equation}
    L(w_m\mid \theta^\star_{m,m+1})=L(w_{m+1}\mid\theta^\star_{m,m+1}).
    \end{equation}
    Note that a marginal change in $w_m$ only changes the neighboring thresholds $\theta^\star_{m-1,m}$ and $\theta^\star_{m,m+1}$, not the others. And if one of these thresholds is a corner, i.e. $\{0,1\}$, then it does not change.
    
    Consider the following partial derivative:
        \begin{multline}
        \frac{\partial \mathcal{L}(\bar{w})}{\partial w_m}=\int_{\theta^\star_{m-1,m}}^{\theta^\star_{m,m+1}}\frac{\partial L(w_m \mid \theta)}{\partial w_m}\pi(\theta)d\theta\\
        +\frac{\partial \theta^\star_{m-1,m}}{\partial w_m}\Big[
        L(w_{m-1} \mid\theta^\star_{m-1,m})-L(w_m\mid\theta^\star_{m-1,m})\Big] \\ \times \pi(\theta^\star_{m-1,m})\\
        +\frac{\partial \theta^\star_{m,m+1}}{\partial w_m}\Big[
        L(w_m \mid \theta^\star_{m,m+1})-L(w_{m+1} \mid \theta^\star_{m,m+1})\Big] \\\times\pi(\theta^\star_{m,m+1}).
    \end{multline}

    The two last terms are zero because either $\theta^*$ is a corner (hence it does not change), or $\theta^\star$ is an interior point and by definition $L(w_{m} \mid \theta^\star_{m,m+1})-L(w_{m+1} \mid \theta^\star_{m,m+1})=0$. Therefore, the first order conditions simplifies to
       \begin{eqnarray*}
        \int_{\theta^\star_{m-1,m}}^{\theta^\star_{m,m+1}}\frac{\partial L(w_m \mid \theta)}{\partial w_m}\pi(\theta)d\theta=0.
    \end{eqnarray*}
    This condition is equivalent to the one in \cref{lem:sender}, and it requires that
       \begin{equation}
        w_m = \mathbf{E}\Big[\Theta \mid \theta^\star_{m-1,m}<\Theta<\theta^\star_{m,m+1}\Big].
    \end{equation}
    The conditions defining $w_m$ and $\theta^\star$ are then the same as the conditions for a BNE. Therefore, the weight vector $\bar{w}$ that minimizes the expected loss is a BNE.
\end{proof}

\section{\textcolor{black}{Numerical Example}}

\begin{figure}[ht!]
    \centering
    \includegraphics[width=\linewidth]{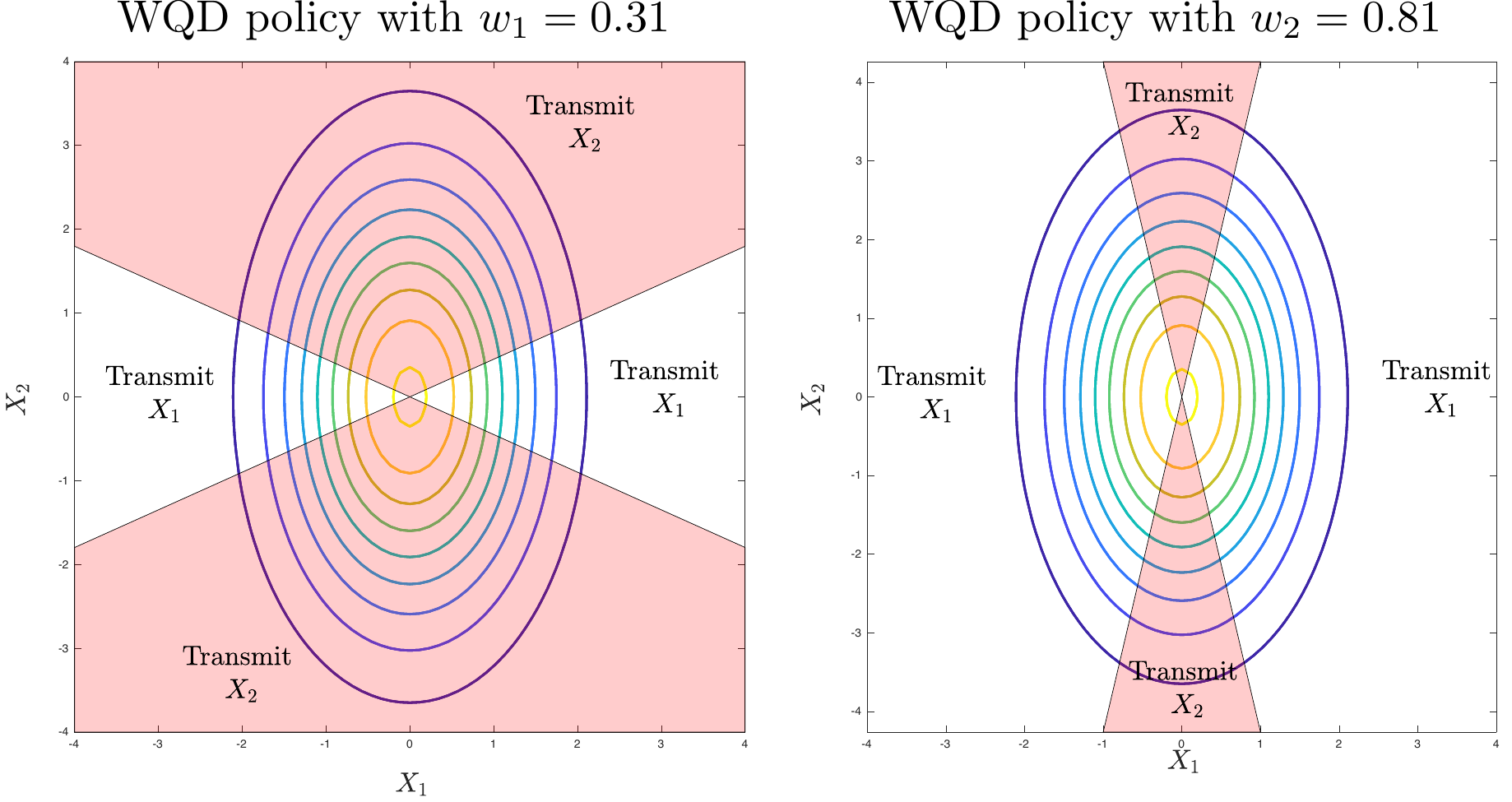}
    \caption{Equilibrium WQD policies used by the Platform in the example.}
    \label{fig:WQD_Example}
\end{figure}

\begin{figure}[ht!]
    \centering
    \includegraphics[width=0.8\linewidth]{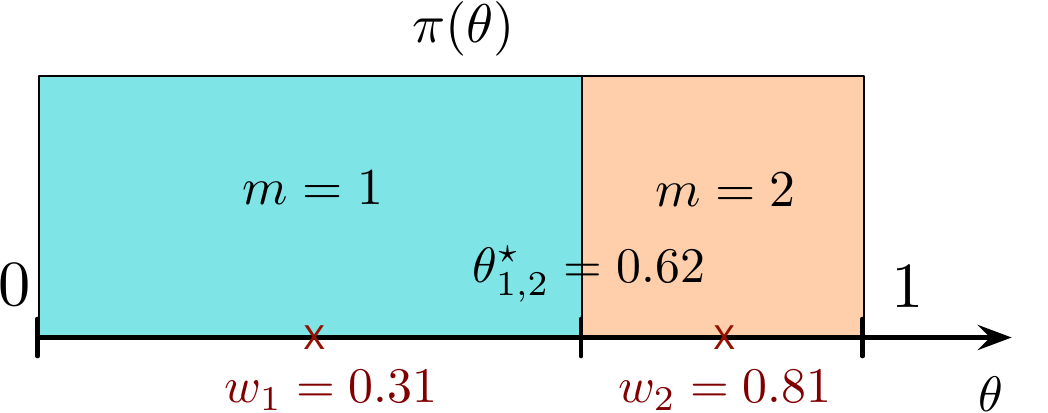}
    \caption{Equilibrium MPPD policies used by the Receiver in the example.}
    \label{fig:MDDP_Example}
\end{figure}

Assume $X_1\sim\mathcal{N}(0,1)$, $X_2\sim\mathcal{N}(0,2)$, $\theta\sim \mathcal{U}[0,1]$, and the receiver can use two messages $n=2$. We \textcolor{black}{use the definition of the loss function in Eq. \eqref{eq:Loss}, together with equilibrium conditions given by equations \eqref{eq.C1} and  \eqref{eq.C1} to} numerically find the following equilibrium. The interior threshold is $\theta^\star_{1,2}\approx0.62$, so that all agents with preference $\theta<\theta^\star_{1,2}$ send message $m=1$ and all agents  $\theta>\theta^\star_{1,2}$ send message $m=2$. The platform applies weight $w_1=\theta^\star_{1,2}/2\approx0.31$ if it receives a message $m=1$, and $w_2=(1+\theta^\star_{1,2})/2\approx0.81$ if it receives a message $m=2$.
The resulting equilibrium policies are depicted in \cref{fig:WQD_Example} for the platform, and in \cref{fig:MDDP_Example} for the receiver.

\section{Conclusions and Future work}

We have introduced a new model of strategic information disclosure between users and a content distribution platform under asymmetric communication constraints, with applications to social media. In our model, the user is characterized by a random private preference variable, initially unknown to the platform, and communicated via a cheap-talk message from a finite discrete alphabet. The platform uses this message to estimate the user's preference and \textcolor{black}{customize its policy governing which information it will transmit to the user,}   
minimizing a weighted quadratic loss function. We prove the existence of a structured Bayesian Nash equilibrium (BNE) pair of policies for both the user and the platform, reminiscent of quantization.

Several directions for future research include incorporating a communication cost for the user and characterizing equilibrium strategies in which the user has an incentive not to transmit a message to the platform. Another extension is to analyze the cost of privacy for receivers who wish to hide their preferences from the platform -- a situation often faced by users interested in content of political nature. Finally, investigating the effects of (possibly intentional) noisy communication in the channel from the user to the platform and its impact on the resulting BNE strategies offers a promising direction.

\bibliography{ref}

\begin{thebibliography}{10}

\bibitem{allon2019information}
G.~Allon, K.~Drakopoulos, and V.~Manshadi, ``Information inundation on
  platforms and implications,'' in {\em Proceedings of the ACM Conference on
  Economics and Computation}, pp.~555--556, 2019.

\bibitem{candogan2020optimal}
O.~Candogan and K.~Drakopoulos, ``Optimal signaling of content accuracy:
  Engagement vs. misinformation,'' {\em Operations Research}, vol.~68, no.~2,
  pp.~497--515, 2020.

\bibitem{milgrom1981good}
P.~R. Milgrom, ``Good news and bad news: Representation theorems and
  applications,'' {\em The Bell Journal of Economics}, pp.~380--391, 1981.

\bibitem{crawford1982strategic}
V.~P. Crawford and J.~Sobel, ``Strategic information transmission,'' {\em
  Econometrica: Journal of the Econometric Society}, pp.~1431--1451, 1982.

\bibitem{Camara:2024}
O.~C{\^a}mara, J.~G. Matsusaka, and C.~Shu, ``Shareholder democracy and the
  market for voting advice,'' {\em Working Paper}, 2024.

\bibitem{Akyol:2017}
E.~Akyol, C.~Langbort, and T.~Ba{\c s}ar, ``Information-theoretic approach to
  strategic communication as a hierarchical game,'' {\em Proceedings of the
  IEEE}, vol.~105, no.~2, pp.~205--218, 2017.

\bibitem{farokhi2016estimation}
F.~Farokhi, A.~M. Teixeira, and C.~Langbort, ``Estimation with strategic
  sensors,'' {\em IEEE Transactions on Automatic Control}, vol.~62, no.~2,
  pp.~724--739, 2016.

\bibitem{Vasconcelos:2020}
M.~M. Vasconcelos and U.~Mitra, ``Observation-driven scheduling for remote
  estimation of two gaussian random variables,'' {\em IEEE Transactions on
  Control of Network Systems}, vol.~7, no.~1, pp.~232--244, 2020.

\bibitem{Vasconcelos:2020b}
M.~M. Vasconcelos, M.~Gagrani, A.~Nayyar, and U.~Mitra, ``Optimal scheduling
  strategy for networked estimation with energy harvesting,'' {\em IEEE
  Transactions on Control of Network Systems}, vol.~7, no.~4, pp.~1723--1735,
  2020.

\bibitem{Vasconcelos:2021}
M.~M. Vasconcelos and U.~Mitra, ``Data-driven sensor scheduling for remote
  estimation in wireless networks,'' {\em IEEE Transactions on Control of
  Network Systems}, vol.~8, no.~2, pp.~725--737, 2021.

\bibitem{vasconcelos2024neuroscheduling}
M.~M. Vasconcelos and Y.~Zhang, ``Neuroscheduling for remote estimation,'' in
  {\em Asilomar Conference on Signals, Systems, and Computers}, 2024.

\bibitem{gray2006quantization}
R.~M. Gray, ``Quantization in task-driven sensing and distributed processing,''
  in {\em IEEE International Conference on Acoustics Speech and Signal
  Processing Proceedings}, vol.~5, IEEE, 2006.

\bibitem{Velicheti:2023}
R.~K. Velicheti, M.~Bastopcu, and T.~Ba{\c s}ar, ``Strategic information design
  in quadratic multidimensional persuasion games with two senders,'' in {\em
  American Control Conference (ACC)}, pp.~1716--1722, 2023.

\bibitem{velicheti2023value}
R.~K. Velicheti, M.~Bastopcu, and T.~Ba{\c{s}}ar, ``Value of information in
  games with multiple strategic information providers,'' {\em arXiv preprint
  arXiv:2306.14886}, 2023.

\bibitem{Velicheti:2024}
R.~K. Velicheti, M.~Bastopcu, S.~R. Etesami, and T.~Ba{\c s}ar, ``Learning how
  to strategically disclose information,'' in {\em American Control Conference
  (ACC)}, pp.~1604--1609, 2024.

\bibitem{ALONSO2016672}
R.~Alonso and O.~C{\^a}mara, ``Bayesian persuasion with heterogeneous priors,''
  {\em Journal of Economic Theory}, vol.~165, pp.~672--706, 2016.

\bibitem{Massicot:2023}
O.~Massicot and C.~Langbort, ``Almost-{B}ayesian quadratic persuasion with a
  scalar prior,'' in {\em 62nd IEEE Conference on Decision and Control (CDC)},
  pp.~228--234, 2023.

\bibitem{saritacs2016quadratic}
S.~Sar{\i}ta{\c{s}}, S.~Y{\"u}ksel, and S.~Gezici, ``Quadratic
  multi-dimensional signaling games and affine equilibria,'' {\em IEEE
  Transactions on Automatic Control}, vol.~62, no.~2, pp.~605--619, 2016.

\bibitem{Hebbar:2022}
V.~Hebbar and C.~Langbort, ``On the role of social identity in the market for
  (mis)information,'' in {\em IEEE 61st Conference on Decision and Control
  (CDC)}, pp.~2512--2518, 2022.

\end{thebibliography}
\bibliographystyle{ieeetr}

\end{document}